\documentclass[
aps,
prb,
amsmath,amssymb,
twocolumn,
superscriptaddress,
showkeys,
showpacs,
nofootinbib,
]{revtex4-1}

\usepackage{bm}
\usepackage{enumitem}
\usepackage{graphicx}
\usepackage[svgnames]{xcolor}
\usepackage{siunitx}
\usepackage[caption=false]{subfig}
\usepackage[colorlinks=true, citecolor=Blue, linkcolor=Blue, urlcolor=Blue]{hyperref}

\usepackage{color,soul}

\usepackage{etoolbox}
\makeatletter
\appto\abstract{
  \let\latexlist\list
  \def\list{\edef\keeprightskip{\the\rightskip}\latexlist}
  \patchcmd\latexlist{\ignorespaces}{\rightskip\keeprightskip\ignorespaces}{}{}
}
\makeatother

\newcommand{\Tc}{T_{\mathrm{c}}}
\newcommand{\NF}{N_{\mathrm{F}}}
\newcommand{\thetaF}{\theta_{\mathrm{F}}}
\newcommand{\kB}{k_{\mathrm{B}}}
\newcommand{\vF}{\mathbf{v}_{\mathrm{F}}}
\newcommand{\pF}{\mathbf{p}_{\mathrm{F}}}
\newcommand{\ps}{\mathbf{p}_{\mathrm{s}}}
\newcommand{\pdef}{2\pi \kB\Tc/v_{\mathrm{F}}}

\begin{document}

\title{Spontaneous symmetry-breaking at surfaces of $d$-wave superconductors:\\ influence of geometry and surface ruggedness}

\author{P. Holmvall}
 \email[]{holmvall@chalmers.se}
 \affiliation{Department of Microtechnology and Nanoscience - MC2, Chalmers University of Technology, SE-41296 G{\"o}teborg, Sweden}
 \author{A. B. Vorontsov}
  \affiliation{Department of Physics, Montana State University, Montana 59717, USA}
\author{M. Fogelstr{\"o}m}
 \affiliation{Department of Microtechnology and Nanoscience - MC2, Chalmers University of Technology, SE-41296 G{\"o}teborg, Sweden}
\author{T. L{\"o}fwander}
 \affiliation{Department of Microtechnology and Nanoscience - MC2, Chalmers University of Technology, SE-41296 G{\"o}teborg, Sweden}

\date{\today}

\begin{abstract}

Surfaces of $d$-wave superconductors may host a substantial density of zero-energy Andreev states. The zero-energy flat band appears due to a topological constraint, but comes with a cost in free energy. We have recently found that an adjustment of the surface states can drive a phase transition into a phase with finite superflow that breaks time-reversal symmetry and translational symmetry along the surface. The associated Doppler shifts of Andreev states to finite energies lower the free energy. Direct experimental verification of such a phase is still technically difficult and controversial, however. To aid further experimental efforts, we use the quasiclassical theory of superconductivity to investigate how the realization and the observability of such a phase are influenced by sample geometry and surface ruggedness. Phase diagrams are produced for relevant geometric parameters. In particular, critical sizes and shapes are identified, providing quantitative guidelines for sample fabrication in the experimental hunt for symmetry-breaking phases.
\end{abstract}

\maketitle



\section{\label{sec:introduction}Introduction}
\begin{figure}[b!]
\includegraphics[width=1.0\columnwidth]{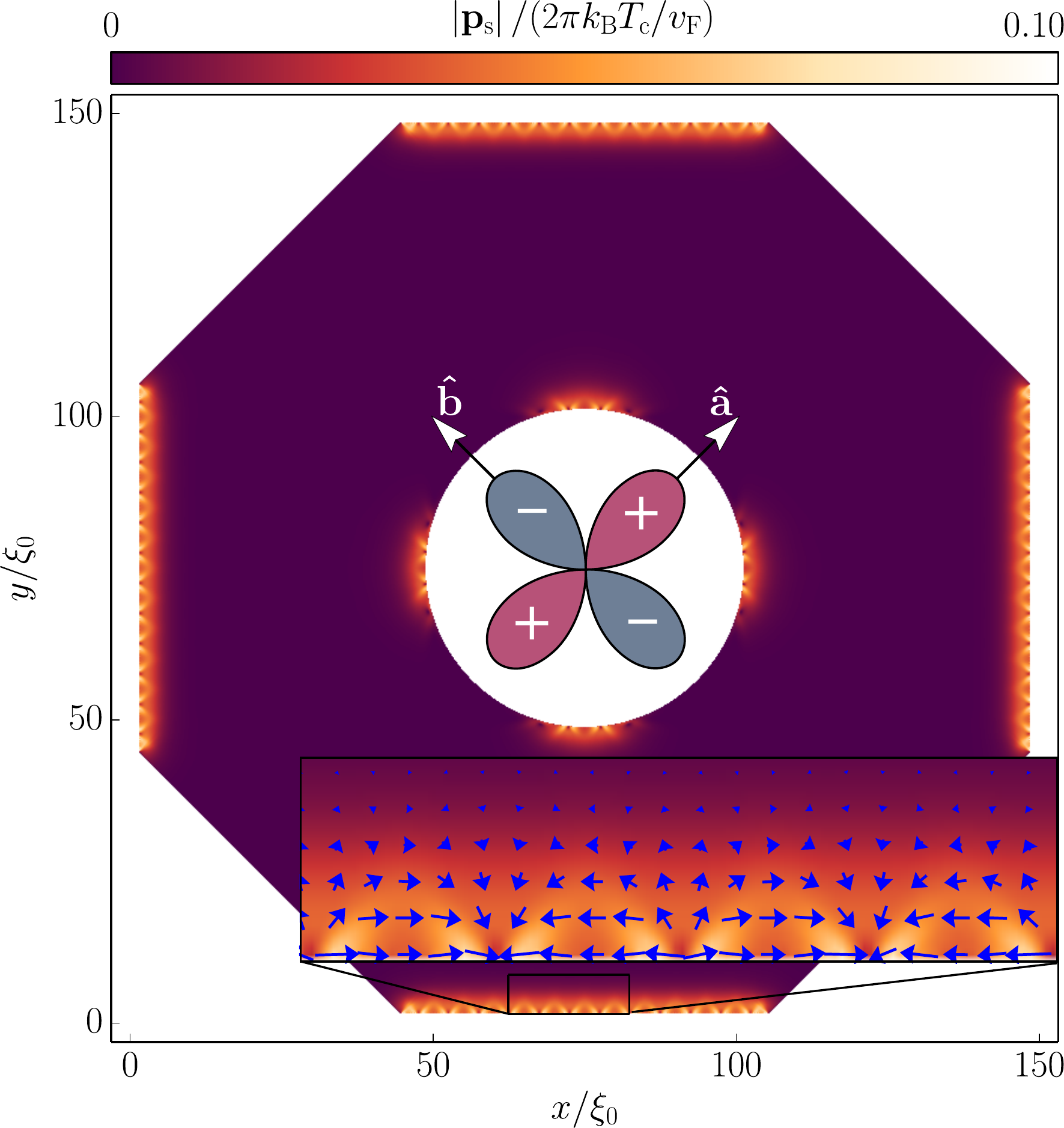}
\caption{A $d$-wave superconducting grain at temperature $T=0.1T_{\mathrm{c}}$ with spontaneous superflow (colors) that spontaneously break translational (along individual surface segments) and time-reversal symmetries. The  Andreev states exist only at the pair-breaking edges, that for this sample geometry occurs along the nodal directions. There is no superflow at surfaces along the lobal directions, since those surfaces have no Andreev states. The inset shows the vector field $\mathbf{p}_{\mathrm{s}}$ (superfluid momentum) with a periodic structure of topological defects \cite{bib:holmvall_2018_a} in the form of edge sources and sinks.
}
\label{fig:system}
\end{figure}

Quasiparticle scattering at interfaces and inhomogeneities of unconventional superconductors leads to pair-breaking and the formation of Andreev states \cite{bib:buchholtz_zwicknagl_1981,bib:hu_1994,bib:lofwander_shumeiko_wendin_2001}.
In a $d$-wave superconductor, these states form a spin-degenerate flat band at zero energy (midgap) that influences tunneling properties, leading to e.g. zero-bias conductance peaks \cite{bib:tanaka_kashiwaya_1995,bib:fogelstrom_rainer_sauls_1997,bib:kashiwaya_tanaka_2000,bib:sauls_2018}.
Furthermore, the Andreev states are bound within a few coherence lengths of the scattering centers, and might influence the superconducting state as a whole in mesoscopic systems \cite{bib:gustafsson_2013}.
The flat band of zero energy states are enforced by topology \cite{bib:sato_tanaka_yada_yokoyama_2011}, but cost free energy.
There are several suggested mechanisms for shifting the states away from the Fermi energy, and thereby lower the free energy in a phase transition where time-reversal ($\cal{T}$) and possibly more symmetries are broken.
In one scenario, a subdominant attractive pairing channel is assumed to exist \cite{bib:Matsumoto_1995a,bib:Matsumoto_1995b,bib:fogelstrom_rainer_sauls_1997,bib:Sigrist_1998}, for instance $s$-wave.
At a temperature $T^*_s$, that depends on the interaction strength in the subdominant channel,
it then becomes energetically favorable to form a composite order parameter $\Delta_d\pm i \Delta_s$, which breaks ${\cal T}$-symmetry and places the Andreev states at $\pm\Delta_s$.
In a second scenario \cite{bib:Honerkamp_2000,bib:Potter_2014}, the repulsive Coulomb interaction in the system may lead to a spin split of the Andreev states, thereby introducing a magnetic transition at a temperature $T^*_{\mathrm{m}}$. 
In a third scenario there are no additional interaction terms in the Hamiltonian. Instead, the appearance of spontaneous superflow sustain Doppler shifts ($\epsilon\rightarrow\epsilon-\vF\cdot\ps$, where $\vF$ is the Fermi velocity and $\ps$ is the superfluid momentum) of the Andreev states to finite energies.
This has been first shown to be possible at translationally invariant surfaces \cite{bib:higashitani_1997,bib:barash_kalenkov_kurkijarvi_2000,bib:lofwander_shumeiko_wendin_2000}.
In this case, the transition temperature $T^*\sim(\xi_0/\lambda)\Tc\ll \Tc$, where $\Tc$ is the superconducting transition temperature of the $d$-wave superconductor, is very low due to the unfavorable ratio between the superconducting coherence length $\xi_0$ and the penetration depth $\lambda$, which appears as a parameter when screening of the surface magnetic field is taken into account.
In a ribbon geometry \cite{bib:vorontsov_2009,bib:hachiya_aoyama_ikeda_2013,bib:higashitani_miyawaki_2015,bib:miyawaki_higashitani_2015_a,bib:miyawaki_higashitani_2015_b},
Andreev states at the two opposite edges interact and hybridize, which provides additional energy shifts that enhance the transition temperature to $T^*\sim(\xi_0/D)\Tc$, where the ribbon width $D$ satisfies $\xi_0<D\ll\lambda$.
Recently\cite{bib:hakansson_2015,bib:holmvall_2018_a,bib:holmvall_2018_b}, we have shown that allowing also breakdown of translational invariance, a single surface will sustain a superflow profile with a texture (see Fig.~\ref{fig:system}) where the associated magnetic flux is restricted to the coherence length scale, in contrast to the penetration depth scale in the translational invariant case.
Thereby, $T^*\sim 0.18 \Tc$ can be achieved for the ideal case of a maximally pair-breaking specular surface of a clean $d$-wave superconductor. This $T^*$ is relatively high, making this scenario very competitive, as long as $T^*>\{T^*_s,\,T^*_{\mathrm{m}}\}$, which depends on the interaction strengths in a particular superconducting material.

There are many experiments that support the claims of symmetry-broken phases \cite{bib:covington_1997,bib:krishana_1997,bib:dagan_2001,bib:gonnelli_2001,bib:elhalel_2007,bib:gustafsson_2013,bib:watashige_2015,bib:bhattacharyya_2018}. However, there are several other experiments that report no signatures, and in particular, no direct imaging of the currents or magnetic fields that would arise in the different scenarios \cite{bib:he_1998,bib:carmi_2000,bib:kirtley_2006,bib:saadaoui_2011,bib:saadaoui_2013}.
Within the scenario of spontaneous superflow with a texture, the breaking of translational invariance leads to inhomogeneous broadening of surface properties probed experimentally on a long length scale compared with the coherence length. The spontanous currents arrange themselves as small loop currents, where neighboring loops have opposite circulation and magnetic field directions. Given the short length scale and the fact that there is no net current flow or flux, such a phase could easily have escaped observation. Such small fluxes and flows would be very difficult to detect unless using very local probes, e.g. single-spin detectors \cite{bib:rugar_2004}, scanning-tunneling spectroscopy \cite{bib:nishio_2008,bib:cren_2011}, nano-SQUIDs \cite{bib:vasyukov_2013}, magnetometry \cite{bib:szechnyi_2017} and diamond cantilevers \cite{bib:pelliccione_2016,bib:ariyaratne_2018}.

To aid such experimental verification, we study in this paper how the realization and observability of the translational symmetry-breaking phase is influenced by sample geometry and surface roughness. In addition, we suggest indirect observation by e.g. penetration-depth measurements \cite{bib:walter_1998}
and nanocalorimetry \cite{bib:diao_rydh_2016,bib:willa_rydh_2017}.

\section{\label{sec:methods}Methods}
We study two-dimensional superconducting grains of various geometries and sizes, with an anisotropic order parameter. In particular, we consider $d$-wave superconductivity with a cylindrically symmetric Fermi surface (see Fig.~\ref{fig:system}), but other order parameters that enable surface Andreev bound states are also of relevance (e.g. polar $p$-wave superconductors \cite{bib:suzuki_asano_2014,bib:dimitriev_2015,bib:zhelev_2016,bib:etter_bouhon_sigrist_2018,bib:miyawaki_higashitani_2018}).
In the present system, the angle between the sample interface and the crystal $ab$-axes (hence the $d$-wave order parameter lobes) directly influences the spectral weight of midgap Andreev states. The grains are assumed to be in vacuum and equilibrium, with spin degeneracy and negligible spin-orbit coupling. Furthermore, the grains are assumed to be clean with perfectly specular interfaces, but effects of disorder and diffuse scattering are discussed.

We utilize the quasiclassical theory of super\-conductivity \cite{bib:eilenberger_1968,bib:larkin_ovchinnikov_1969,bib:shelankov_1985,bib:serene_rainer_1983,bib:eschrig_heym_rainer_1994,bib:eschrig_rainer_sauls_1999,bib:sauls_eschrig_2009}, in which the Green's function $\hat{g}(\pF, \mathbf{R}; z)$ governs quasiparticle and pair propagation through the Eilenberger equation
\begin{equation}
\label{eq:eilenberger}
i\hbar \vF \cdot \bm{\nabla}_{R}\,\hat{g} + \left[\hat{\tau}_3\left(z + \vF \cdot \frac{e}{c}\mathbf{A}\right)-\hat{h},\hat{g}\right] = \hat{0},
\end{equation}
with the normalization condition
\begin{equation}
\label{eq:normalization}
\hat{g}^2 = -\pi^2\hat{1}.
\end{equation}
Here, $\pF$ is the quasiparticle momentum at the Fermi-surface, $\mathbf{R}$ the center-of-mass coordinate, $z$ the energy, $\hbar$ the reduced Planck constant, $\vF$ the Fermi velocity, $e$ the elementary charge, $c$ the speed of light, $\mathbf{A}$ the electromagnetic gauge field, and $\hat{\tau}_3$ the third Pauli matrix, where the hat symbol denotes Nambu (electron-hole) space. The self-energies $\hat{h}$ are expressed in terms of the superconducting order parameter $\Delta$,
\begin{equation}
\label{eq:self_energies}
\hat{h} = 
\left(
\begin{array}{cc}
{0} & {\Delta}\\
{\tilde{\Delta}} & {0}
\end{array}\right),
\end{equation}
and the quasiclassical Green's function is described in terms of the quasiparticle and pair-propagators $g$ and $f$, respectively,
\begin{equation}
\label{eq:g}
\hat{g} = 
\left(
\begin{array}{cc}
{g} & {f}\\
-{\tilde{f}} & {\tilde{g}}
\end{array}\right).
\end{equation}
The tilde symbol denotes particle-hole conjugation
\begin{equation}
\label{eq:tilde_operator}
\tilde{\alpha}(\pF,\mathbf{R};z) = \alpha^*(-\pF,\mathbf{R};-z^*).
\end{equation}
To solve the Eilenberger equation, the Riccati formalism is used \cite{bib:nagato_1993,bib:schopohl_maki_1995,bib:schopohl_1998,bib:eschrig_sauls_rainer_1999,bib:eschrig_2000,bib:vorontsov_sauls_2003,bib:eschrig_2009}, in which the quasiclassical Green's function is parametrized in terms of two particle-hole coherence functions $\gamma(\pF, \mathbf{R}; z)$ and $\tilde{\gamma}(\pF, \mathbf{R}; z)$
\begin{equation}
\label{eq:g_of_gamma}
\hat{g} = -\frac{i\pi}{1+\gamma\tilde{\gamma}}
\left(
\begin{array}{cc}
{1-\gamma\tilde{\gamma}} & {2\gamma}\\
{2\tilde{\gamma}} & {-1+\gamma\tilde{\gamma}}
\end{array}\right),
\end{equation}
yielding two Riccati equations
\begin{eqnarray}
\label{eq:riccati_gamma}
\left(i\hbar\vF\cdot\bm{\nabla}_{R} + 2z +2\frac{e}{c}\vF\cdot\mathbf{A}\right)\gamma & = & -\tilde{\Delta}\gamma^2 - \Delta,\\
\label{eq:riccati_gamma_tilde}
\left(i\hbar\vF\cdot\bm{\nabla}_{R} - 2z -2\frac{e}{c}\vF\cdot\mathbf{A}\right)\tilde{\gamma} & = & -{\Delta}\tilde{\gamma}^2 - \tilde{\Delta}.\end{eqnarray}
In this paper, a pure $d$-wave order parameter is assumed
\begin{eqnarray}
\label{eq:d_wave}
\Delta(\pF,\mathbf{R}) & = & \Delta_d(\mathbf{R})\eta_d(\theta_{\mathrm{F}}),\\
\label{eq:d_wave_basis}
\eta_d(\theta_{\mathrm{F}}) & = & \sqrt{2}\cos(2\theta_{\mathrm{F}}),
\end{eqnarray}
where $\eta_d$ is the $d$-wave basis function, and $\theta_{\mathrm{F}}$ is the angle between the Fermi momentum and the crystal $\mathbf{\hat{a}}$-axis. The order parameter is solved self-consistently from the gap equation with the Matsubara technique
\begin{equation}
\label{eq:gap_equation}
\Delta_d(\mathbf{R}) = \lambda_d\NF\kB T\sum_{|\epsilon_{m}|\leq\Omega_{\mathrm{c}}}\int\frac{d\theta_{\mathrm{F}}}{2\pi}\eta_d^*(\theta_{\mathrm{F}})f(\pF,\mathbf{R};i\epsilon_m),
\end{equation}
where $\lambda_d$ is the pairing interaction, $\kB$ the Boltzmann constant, $\epsilon_m$ the Matsubara energy, $\Omega_{\mathrm{c}}$ is a cutoff energy and $\NF$ is the normal-state density of states at the Fermi surface (per spin).

This theoretical formalism is implemented numerically to run on graphics processing units, where the above equations of motion are solved in parallel over different degrees of freedom, until self-consistency is achieved, see Refs.~\cite{bib:hakansson_2015,bib:hakansson_lic_2015,bib:holmvall_lic_2017,bib:holmvall_2018_a} for more details. Finally, various quantities are calculated, e.g. the gauge-invariant superfluid momentum $\ps$ which we have identified \cite{bib:holmvall_2018_a} as the order parameter of the symmetry-broken phase
\begin{equation}
\label{eq:ps}
\ps(\mathbf{R})=\frac{\hbar}{2}\nabla\chi(\mathbf{R})-\frac{e}{c}\mathbf{A}(\mathbf{R}),
\end{equation}
where $\chi$ is the superconducting phase.
The local density of states (DOS) at energy $\epsilon$ is calculated as a Fermi-surface average
\begin{equation}
\label{eq:ldos}
N(\mathbf{R}; \epsilon) = -\frac{\NF}{\pi}\int\frac{d\theta_{\mathrm{F}}}{2\pi}\operatorname{Im}\left[ g(\pF,\mathbf{R};\epsilon+i\delta)\right],
\end{equation}
where $\delta\rightarrow 0^+$ guarantees a retarded Green's function.
The current density is calculated according to
\begin{equation}
\label{eq:current}
\mathbf{j}(\mathbf{R}) = 2\pi e\NF\kB T\sum_{\epsilon_m}\int\frac{d\theta_{\mathrm{F}}}{2\pi}\mathbf{v}_{\mathrm{F}}g(\pF,\mathbf{R};i\epsilon_m).
\end{equation}
In absence of impurity scattering, the free-energy difference between the superconducting and the normal state, at temperature $T$, may be calculated as \cite{bib:eilenberger_1968}
\begin{eqnarray}
\label{eq:free_energy}
\Omega(T) & = & \int d\mathbf{R}\Bigg\{\frac{\mathbf{B}^2(\mathbf{R})}{8\pi} + |\Delta(\mathbf{R})|^2\NF\ln\frac{T}{\Tc} + \nonumber\\
& & 2\pi \NF\kB T\sum_{\epsilon_m>0}\Bigg[\frac{|\Delta(\mathbf{R})|^2}{\epsilon_m} + i\mathcal{I}(\mathbf{R},\epsilon_m)\Bigg]\Bigg\},\\
\mathcal{I}(\mathbf{R}) & = & \int\frac{d\theta_{\mathrm{F}}}{2\pi}\Big[\tilde{\Delta}(\pF,\mathbf{R})\gamma(\pF,\mathbf{R};i\epsilon_m) - \nonumber\\
& & \Delta(\pF,\mathbf{R})\tilde{\gamma}(\pF,\mathbf{R};i\epsilon_m)\Big],
\end{eqnarray}
where $\mathbf{B}$ is the induced magnetic field and $\Tc$ the superconducting transition temperature. The heat capacity is obtained from the free energy according to
\begin{equation}
\label{eq:heat_capacity}
C(T) = -T\frac{\partial^2\Omega(T)}{\partial T^2}.
\end{equation}

\section{\label{sec:results}Results and discussion}
We start by varying the angle between the interface and the crystal $\mathbf{\hat{a}}$-axis (Sec.~\ref{sec:results:angle}), thus controlling the pair-breaking effect. We then vary the area of the grain to study finite-size effects (Sec.~\ref{sec:results:area}). Critical angles and areas are identified. These results are used to analyze superconducting grains of various shapes and different degrees of surface roughness (Sec.~\ref{sec:results:roughness}). We will limit ourselves to mesoscopic roughness (see below), and we will consider clean superconductors. It is known (see e.g. the review in Ref.~\cite{bib:lofwander_shumeiko_wendin_2001}) that impurities will broaden the Andreev state peak around zero energy, as will a finite transparency interface to other materials, for instance normal metals used in tunnelling experiments. These broadening effects may suppress $T^*$, but we leave their effects for future studies.

\subsection{\label{sec:results:angle}Critical interface angle}
\begin{figure}[b]
\includegraphics[width=\columnwidth]{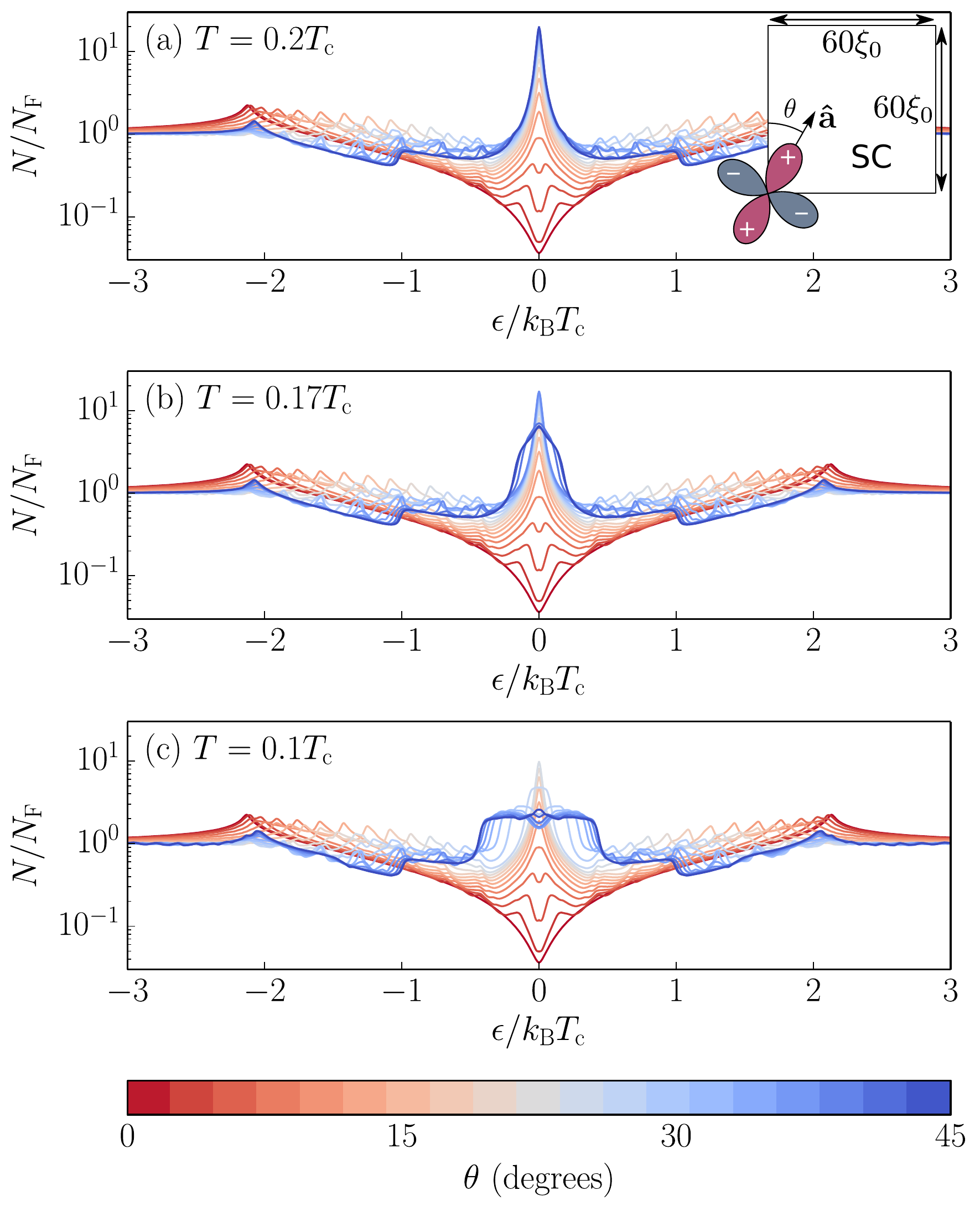}
\caption{The surface density of states averaged over one side of a $60\xi_0 \times 60\xi_0$ square grain with $\operatorname{Im}{z} = \delta = 0.02\kB\Tc$. Different curves correspond to different values of $\theta$, as indicated by colors. (a) Above the transition, the peak is narrow (note the logarithmic scale on the ordinate). The steps at $\epsilon \approx 1\kB\Tc \approx \Delta_0/2$ come from the features in the DOS at the square corners, with $\Delta_0\approx2.14\kB\Tc$ being the bulk gap. (b,c) As the temperature is lowered, the midgap states are broadened due to the presence of spontaneous superflow. }
\label{fig:results:SDOS}
\end{figure}

\begin{figure}[b]
\includegraphics[width=\columnwidth]{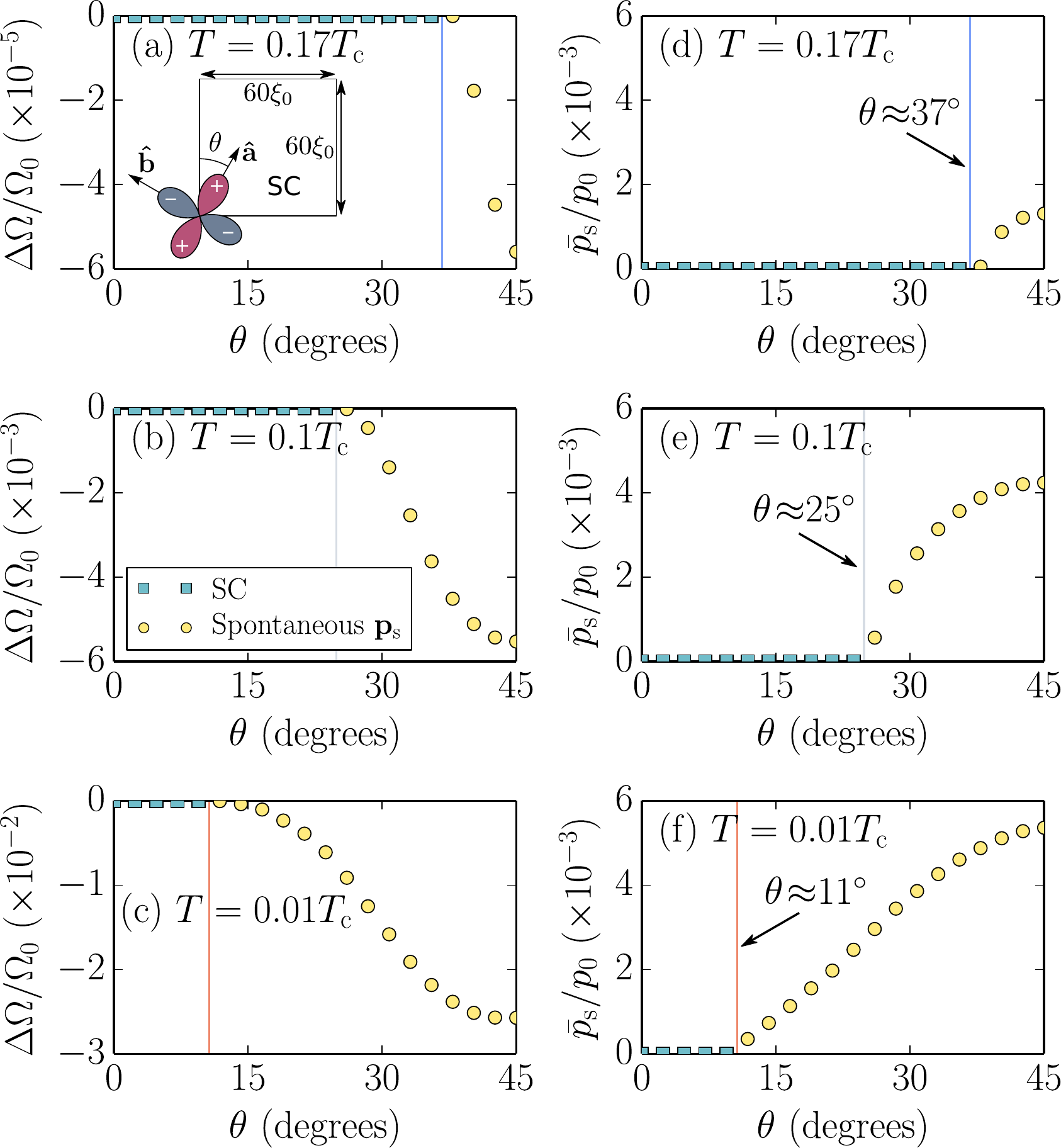}
\caption{ (a)--(c) Free-energy difference between phases with and without spontaneous superflow, and (d)--(f) sample-averaged magnitude of the spontaneous superfluid momentum. These quantities are plotted versus the angle $\theta$ between the grain edges and the crystal $\mathbf{\hat{a}}$-axis in a square grain with area $\mathcal{A} = (60 \xi_0)^2$, as illustrated in the inset in (a). Here, the units are $\Omega_0 \equiv \mathcal{A}\NF\kB^2\Tc^2$ and $p_0 \equiv \pdef$. Note the varying scales in (a)--(c). }
\label{fig:results:free_energy_and_ps}
\end{figure}

\begin{figure}[b]
\includegraphics[width=\columnwidth]{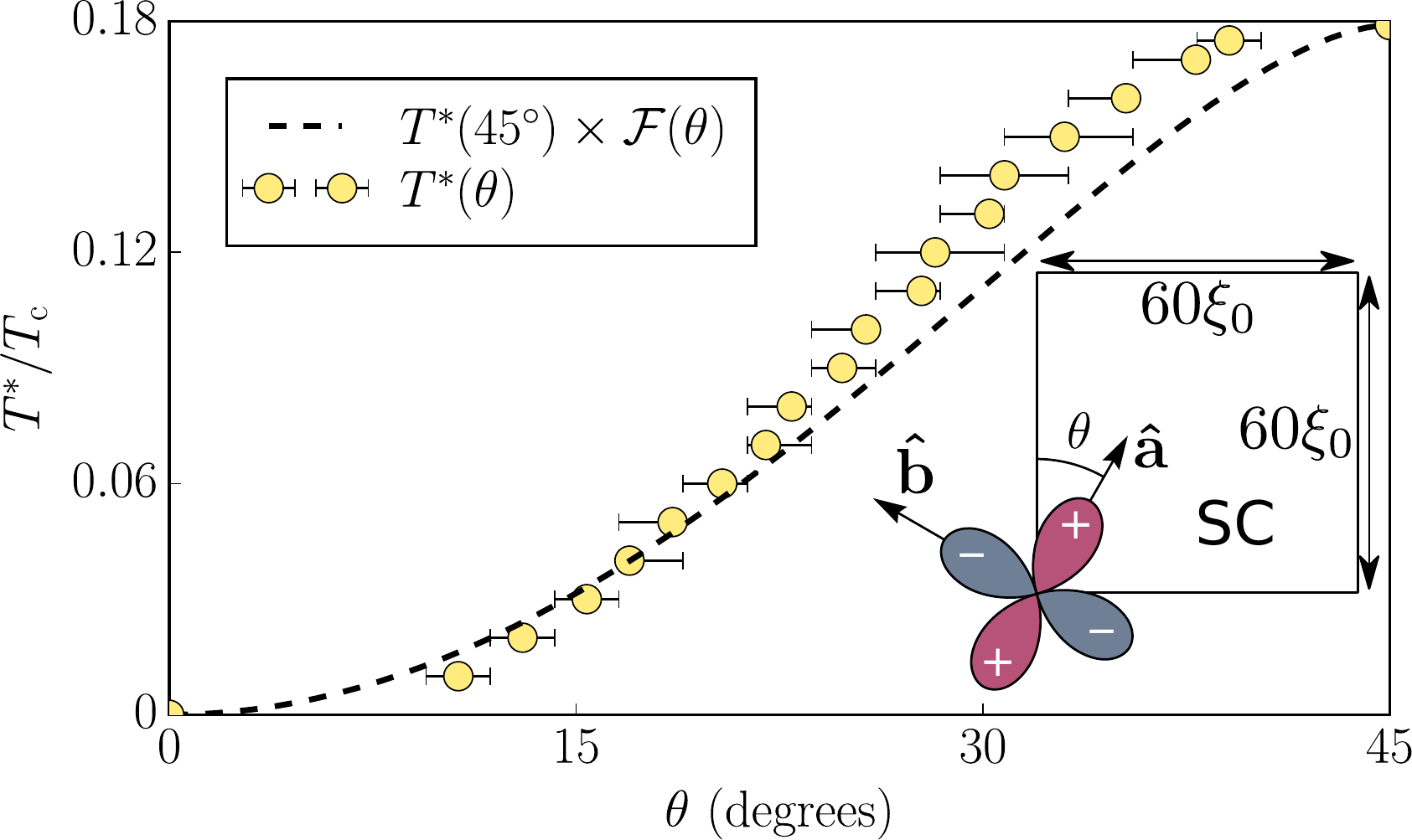}
\caption{ Phase diagram showing the transition temperature $T^{*}$ of the symmetry-broken phase, as a function of the angle $\theta$ between a vacuum-superconductor interface and the $d$-wave crystal $\mathbf{\hat{a}}$-axis, in a grain of area $\mathcal{A} = (60 \xi_0)^2$ (see inset). Error bars denote the uncertainty due to the discrete angular resolution. $T^{*}(\theta)$ is roughly described by the angular dependence of the MGS peak $N_{\mathrm{MGS}}$, denoted $\mathcal{F}(\theta)$ (dashed line), defined in Eq.~(\ref{eq:N_MGS:2}).
}
\label{fig:results:angle_phase_diagram}
\end{figure}

As the angle $\theta$ between a specular $d$-wave interface and the crystal $\mathbf{\hat{a}}$-axis is varied from perfectly aligned ($\theta = 0^{\circ}$) to perfectly misaligned ($\theta = 45^{\circ}$), the surface DOS changes from the typical gapless bulk DOS to one with a large zero-energy peak, as illustrated in Fig.~\ref{fig:results:SDOS} (a).
The states in the peak come from quasiparticles that scatter between directions with a sign change in the order parameter. 
These midgap states (MGS) are thus enforced by the order parameter symmetry, and associated with a significant increase in free energy
and also a suppression of the order parameter at the interface.
As the temperature is lowered, there is a phase transition at $T^{*}$ where superflow appears spontaneously.
The energy is lowered by Doppler shifting the zero-energy states to finite energies, as seen in Fig.~\ref{fig:results:SDOS} (b). 
The magnitude of superflow, the $\ps$ field seen in Fig.~\ref{fig:system}, increases as the temperature is lowered \cite{bib:holmvall_2018_a}, as does then the Doppler shift and the energy gain as well, as seen by comparing Figs.~\ref{fig:results:SDOS} (b) and (c).
Figures~\ref{fig:results:free_energy_and_ps} (a)--(c) show the free-energy gain $\Delta \Omega$ versus $\theta$, defined as
\begin{equation}
\label{eq:free_energy_difference}
\Delta \Omega = \Omega_{\mathrm{S}} - \Omega_{\mathrm{ms}},
\end{equation}
where $\Omega_S$ and $\Omega_{\mathrm{ms}}$ are the free energies of the systems with and without spontaneous superflow, respectively. The latter might exhibit a higher a free energy and is therefore referred to as a metastable state (ms).
Figures~\ref{fig:results:free_energy_and_ps} (d)--(f) show the sample-averaged magnitude of the superfluid momentum $\bar{p}_{\mathrm{s}}$ versus $\theta$, defined as
\begin{equation}
\label{eq:ps_averaged}
\bar{p}_{\mathrm{s}} = \frac{1}{\mathcal{A}}\int d\mathbf{R} |\ps(\mathbf{R})|,
\end{equation}
where $\mathcal{A}$ is the sample area.
From these figures, it is possible to identify a lowering of the free energy with $\ps \neq 0$. A critical phase transition temperature $T^*(\theta)$, 
defined as the temperature where $\bar{p}_{\mathrm{s}}$ becomes finite, is plotted in a phase diagram in Fig.~\ref{fig:results:angle_phase_diagram}. Error bars originate from the uncertainty due to the discrete angular resolution.
The transition temperature $T^*$ closely follows the spectral weight of the MGS peak, that can be controlled by various parameters such as surface roughness, or, as in this case, by the interface orientation $\theta$.
This can be shown from a very general argument as follows. 
The gain in free energy due to a small shift of zero-energy states with narrow DOS 
$N_{\mathrm{MGS}}(\epsilon) = N_{\mathrm{bs}}\delta(\epsilon)$ by  
$\Delta \epsilon$ (e.g. $\propto \vF \cdot \ps$ Doppler shift) 
is\cite{bib:vorontsov_2018}
\begin{align}\begin{split}
\Delta F_{\mathrm{b}}(T) = -\int_{-\infty}^{\infty} d\epsilon \; 
& \kB T\ln\left(2\cosh\frac{\epsilon}{2\kB T}\right) \\
& \times \left[N_{\mathrm{bs}}\delta(\epsilon-\Delta\epsilon) - N_{\mathrm{bs}}\delta(\epsilon)\right] \,.
\end{split}\label{eq:free_energy_MGS:1}\end{align}
that for $\Delta\epsilon \ll \kB T$ reduces to
\begin{equation}
\label{eq:free_energy_MGS:2}
\Delta F_{\mathrm{b}}(T) \approx -N_{\mathrm{bs}}\frac{\left(\Delta\epsilon\right)^2}{8\kB T},
\end{equation}
The same spectral shift of the continuum states, however, increases energy, 
also $\propto (\Delta\epsilon)^2$ (e.g. superflow energy $\propto \ps^2$) as  
$\Delta F_{\mathrm{c}}(T) = A(T) (\Delta \epsilon)^2 $, where the parameter $A(T)$ depends on the mechanism of the energy increase, and in principle should take into account the reduction of continuum states by $N_{\mathrm{bs}}$. The instability occurs when their sum is negative 
\begin{align}\begin{split}
\Delta F_{\mathrm{b}}(T^*) + \Delta F_{\mathrm{c}}(T^*) \le 0 
\\ \quad \Rightarrow \quad  
T^* \approx \mathrm{const} \frac{N_\mathrm{bs}}{A(T^*)}
\end{split}\label{eq:free_energy_MGS:3}\end{align}
Assuming that $A(T)$ is relatively insensitive to temperature and to the transfer of spectral weight to bound states, the main effect on the transition temperature is from varying $N_\mathrm{bs}$
\begin{equation}
\label{eq:free_energy_MGS:4}
T^* = \mathrm{const} \cdot N_\mathrm{bs} \,.
\end{equation}
This argument can be further adjusted for broadening of the bound states by impurities for example, and corrected for the continuum reduction $\delta T^*\propto O(N_\mathrm{bs}^2)$.
For the $\theta$-rotation of the crystal axes we can estimate the height of the bound state peak $N_\mathrm{bs}$ analytically.  
Neglecting the order parameter suppression the low-energy Green's function at the surface is 
($|z=\epsilon + i\delta| \ll |\Delta_{\mathrm{in,out}}|$)
\begin{equation}
\label{eq:g_MGS}
g\left(z\right) = \frac{\pi}{z}  \frac{2 |\Delta_{\mathrm{in}} \Delta_{\mathrm{out}}|}{|\Delta_{\mathrm{in}}|+|\Delta_{\mathrm{out}}|} \Theta(-\Delta_{\mathrm{in}} \Delta_{\mathrm{out}}),
\end{equation}
where $\Theta$ is the Heaviside function, and $\Delta_{\mathrm{in/out}}=\Delta(\thetaF),\; \Delta(\pi-\thetaF)$ are the order parameters for incoming and outgoing trajectories, respectively. Averaging the DOS over the Fermi surface, as in Eq.~(\ref{eq:ldos}), we get 
\begin{eqnarray}
\label{eq:N_MGS:1}
N_{\mathrm{MGS}}(\epsilon,\theta)  =  - 2\NF \operatorname{Im} \frac{\Delta_0}{\epsilon +i\delta} \frac{2}{\pi} \mathcal{F}(\theta), \quad \Rightarrow \\
\label{eq:N_MGS:2}
N_\mathrm{bs} \propto \mathcal{F}(\theta)  \equiv  1 - \frac{\cos^22\theta}{\sin 2\theta}\ln\left(\frac{1+\tan\theta}{1-\tan\theta}\right),
\end{eqnarray}
where $\Delta_0$ is the bulk gap amplitude. Scaling of transition temperature by the zero-energy spectral weight $\mathcal{F}(\theta)$ is shown by the dashed line in Fig.~\ref{fig:results:angle_phase_diagram}. It shows a very close relation with the full numerical result, given the roughness of our estimate.

The phase diagram in Fig.~\ref{fig:results:angle_phase_diagram} shows that there is robustness of the symmetry-broken phase against surface disorder at $d$-wave interfaces, and that even completely circular interfaces will host the phase as long as the radius of curvature is large enough, as seen in e.g. Fig.~\ref{fig:system}.

\subsection{\label{sec:results:area}Critical grain area}
The spectral weight of zero-energy states is peaked at the interface, but extends almost $10 \xi_0$ away from it. Square grains with sidelengths smaller than $20 \xi_0$ therefore exhibit pronounced finite-size effects, e.g. suppressed superconductivity and a reduced $T_{\mathrm{c}}$, due to overlapping regions of MGS. In larger systems, the MGS from different interfaces will no longer overlap except in the corners. 
Quantities which are directly tied to the MGS, e.g. $|\ps(\mathbf{R})|$ and $\bm{j}$, are therefore expected to show a saturation for larger grain sizes.

\begin{figure}[b]
\includegraphics[width=\columnwidth]{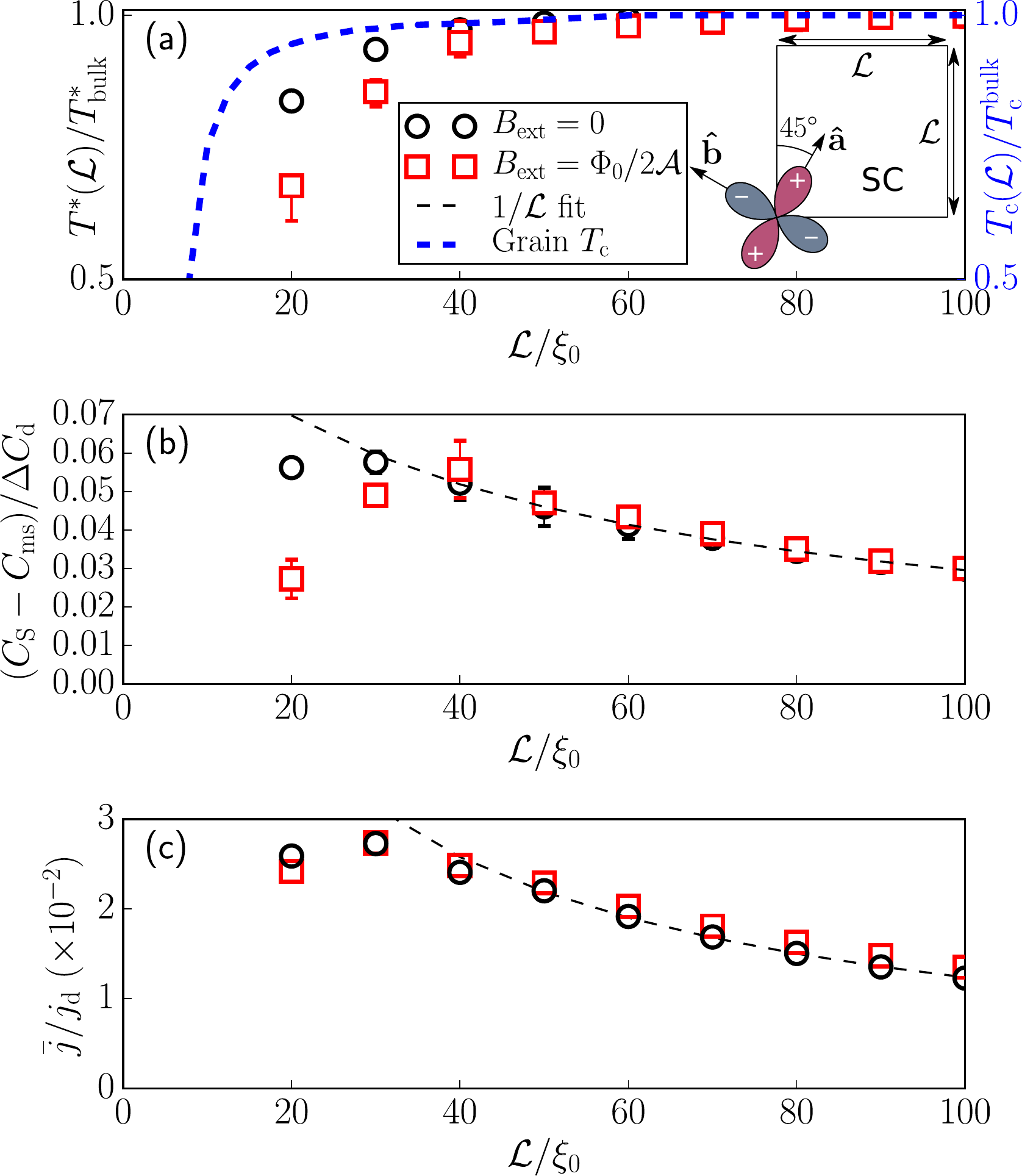}
\caption{ Effect of sample size $\mathcal{L}\times\mathcal{L}$ with maximal pairbreaking edges (see inset) on (a) the transition temperature into the spontaneous superflow phase (circles and squares, left axis) and the superconducting transition temperature of the grain (dashed line, right axis), (b) the heat capacity jump of the spontaneous superflow phase transition, and (c) the sample-averaged current magnitude at $T=0.1T_{\mathrm{c}}$. Error bars denote uncertainty due to discrete resolution in temperature (a), and numerical uncertainty in the heat capacity (b). }
\label{fig:results:size}
\end{figure}

We will now quantify how the sidelength $\mathcal{L}$ of a square grain with maximally pair-breaking interfaces ($\theta = 45^{\circ}$) influences the transition temperature $T^{*}$, the heat capacity jump, as well as the average current magnitude of the symmetry-broken phase. Since the phase under investigation is a second-order phase transition, the transition temperature is appropriately extracted from where there is a discontinuity in the heat capacity \cite{bib:holmvall_2018_a}. Figure \ref{fig:results:size} (a) shows $T^{*}(\mathcal{L})$ with and without an external magnetic field (circles and squares, respectively, left axis), and $\Tc(\mathcal{L})$ of the grain (thick dashed line, right axis).
Here and in the following, the external magnetic field corresponds to half a flux quantum spread across the grain area $B_{\mathrm{ext}} = \Phi_0/2\mathcal{L}^2$, where $\Phi_0 \equiv hc/2|e|$ is the unit of flux quantum. The deviation from $\Tc(\mathcal{L}) = \Tc^{\mathrm{bulk}}$ indicates finite-size effects. Hence, $T^{*}$ decreases with $\mathcal{L}$ due to superconductivity being suppressed in the grain. The suppression is stronger with an external field as the resulting screening currents also suppress superconductivity. As the sidelength increases, the regions of MGS no longer overlap and saturate to fixed sizes and shapes. The transition temperature therefore also saturates to a fixed value. Figure~\ref{fig:results:size} (b) shows how the sample-average heat-capacity jump changes with the sidelength (with and without external field), while Fig.~\ref{fig:results:size} (c) shows the sample-averaged magnitude of the current, defined as
\begin{equation}
\label{eq:j_average}
\bar{j} = \frac{1}{\mathcal{A}}\int d\mathbf{R} |\mathbf{j}(\mathbf{R})|.
\end{equation}
The heat-capacity jump in the bulk normal-superconducting phase transition is given by
\begin{equation}
\label{eq:hc_d}
\Delta C_d = \frac{2\alpha}{3} {\cal A} \kB^2\Tc\NF,
\end{equation}
where $\alpha = 8\pi^2/[7\zeta(3)]$ and $\zeta$ is the Riemann-zeta function. Again, finite-size effects can be seen in Figs.~\ref{fig:results:size} (b) and (c) due to suppression of superconductivity at smaller $\mathcal{L}$.
Furthermore, since the superfluid momentum is directly tied to the MGS, both $\ps$ and $\mathbf{j}$ saturate to fixed profiles at larger $\mathcal{L}$. Sample-averaged quantities, e.g. $\bar{j}$ and $(C_{\mathrm{S}}-C_{\mathrm{ms}})/\Delta C_{\mathrm{d}}$ thus scale as $\mathcal{L}^{-1}$, as evident by the fit. The fit breaks down at the onset of finite-size effects, resulting in a maximum at a finite $\mathcal{L} = \mathcal{L}_{\mathrm{c}} \approx 30\xi_0$.

These results imply that the observability of the phase through sample-averaged observables is maximized at a finite sidelength. This ratio will depend on the shape of the sample, and in particular the angles of the interfaces. Therefore, for e.g. thermodynamic experiments aiming to verify the symmetry-broken phases, it might be advisable to fabricate e.g. thin rectangular grains or square grains of sidelengths $\sim 30\xi_0$, depending on the type of experiment. On the other hand, if the goal is instead to avoid this phase, very small grains with $\mathcal{L}<\mathcal{L}_{\mathrm{c}}$ are advisable.

\begin{figure}[b]
\includegraphics[width=\columnwidth]{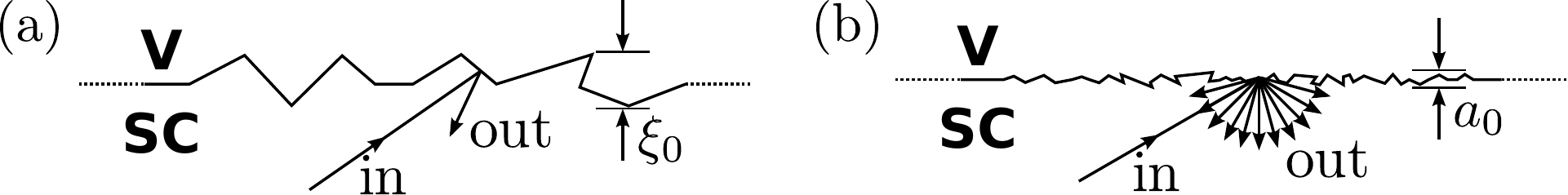}
\caption{ (a) Mesoscopic surface roughness, where the disorder is on the coherence length scale or larger. The roughness is modeled as mesoscopic facets that scatter incoming quasiparticle states specularly. (b) Atomic surface roughness, where the disorder is on the atomic scale, i.e. generally much smaller than the superconducting coherence length, leading to a diffuse scattering of any incoming quasiparticle states. }
\label{fig:surface_roughness}
\end{figure}

\subsection{\label{sec:results:roughness}Surface roughness}
With the quantitative knowledge about how the size and the angle of the pair-breaking interfaces influence the symmetry-broken phase, we will now qualitatively study the effect of surface roughness. There are two well-defined regimes of surface roughness, here referred to as mesoscopic roughness (or ruggedness) and atomic surface roughness, as illustrated in Fig.~\ref{fig:surface_roughness}.
\begin{figure*}[t!]
\includegraphics[width=1.0\linewidth]{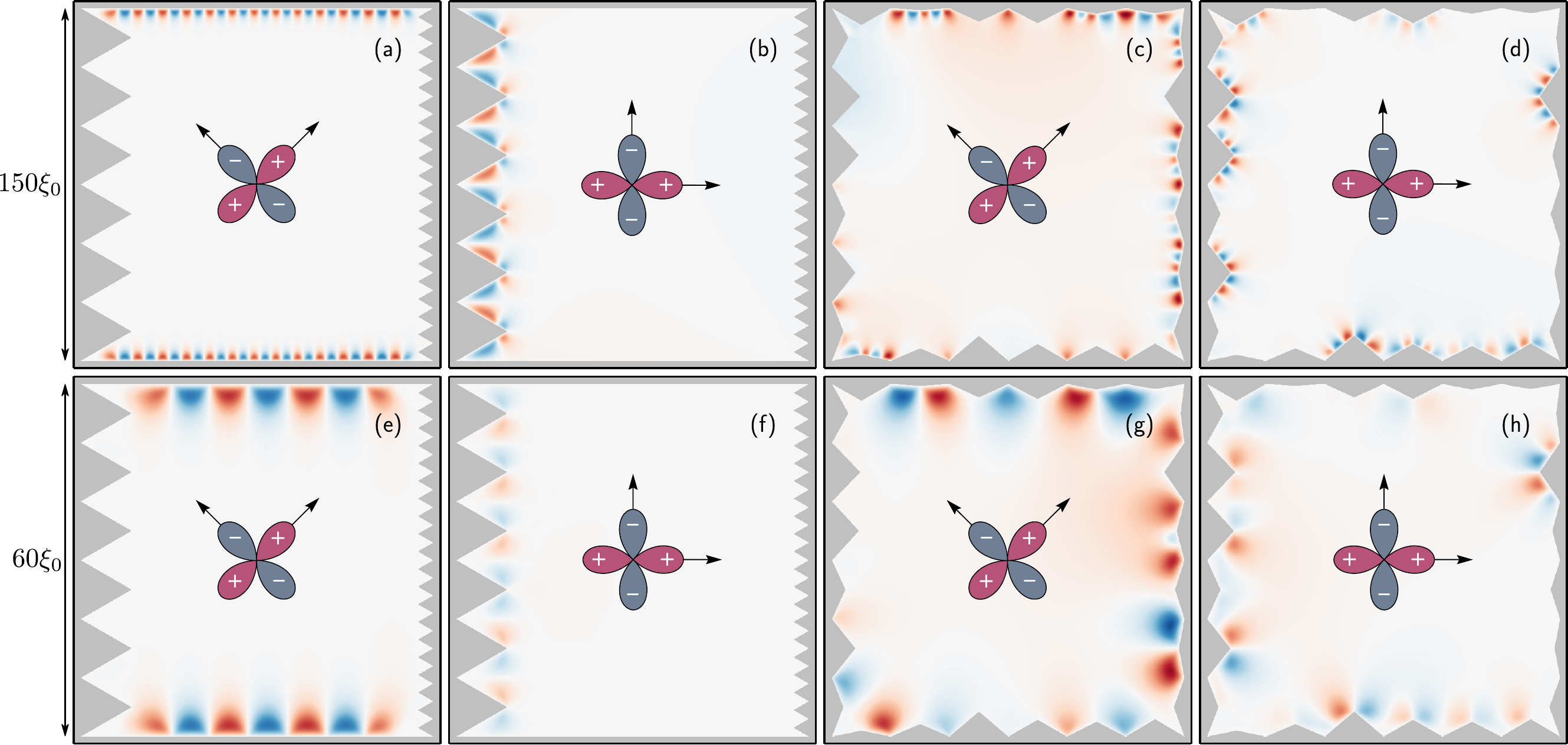}
\caption{ $d$-wave grains with sidelengths (a)--(d) $150\xi_0$ and (e)--(h) $60\xi_0$, and with different degrees of mesoscopic surface roughness (gray boundaries). Colors indicate the magnitude of magnetic fields induced by spontaneous currents, with a maximum/minimum flux density of roughly $\pm 10^{-5}\Phi_0/\xi_0^2$, where $\Phi_0 \equiv hc/2|e|$ is the magnetic flux quantum. }
\label{fig:results:surface_roughness_samples}
\end{figure*}

Mesoscopic surface roughness refers to interfaces with a disorder that is on the order of the superconducting coherence length $\xi_0$ or larger, i.e. mesoscopic facets that scatter specularly. For high-temperature superconductors the coherence length is very short and this kind of ruggedness instead of atomic scale roughness can be a relevant regime. Figure \ref{fig:results:surface_roughness_samples} shows spontaneous magnetic fields caused by spontaneous superflow in square grains with sidelengths of (a)--(d) $150\xi_0$ and (e)--(h) $60\xi_0$, with varying degrees of mesoscopic roughness. It is seen that despite a rugged surface profile, the spontaneous superflow might appear. The two key prerequisites are that the facet angle with respect to the crystal $\mathbf{\hat{a}}$-axis must lie within the critical angle quantified in Fig.~\ref{fig:results:angle_phase_diagram}, and that the area around the facet is large enough to accommodate the superfluid momentum profile. These findings illustrate that the symmetry-broken phase is relatively robust against mesoscopic roughness.

Atomic surface roughness on the other hand refers to surfaces that have a disorder that is on the atomic length scale, e.g. the Bohr radius $a_0$ or the Fermi wavelength $\lambda_{\mathrm{F}}$, which are both generally smaller than the superconducting coherence length. This disorder will lead to diffuse scattering of any incoming quasiparticle state, with a finite probability of backscattering. Hence, while a clean pair-breaking $d$-wave interface will induce a sign change for most quasiparticle scattering trajectories, diffusivity will severely reduce the number of such trajectories and thus also the spectral weight of midgap states. It was previously shown that the symmetry-broken phase in ribbons persisted up to roughly $80\%$ diffusivity \cite{bib:higashitani_miyawaki_2015}. For polar $p$-wave superconductors with the nodal direction along the interface, the order-parameter sign change accompanies all scattering trajectories independent of conservation of $p_{\parallel}$ (in contrast to $d$-wave superconductors). The zero-energy states in such a $p$-wave superconductor will thus be completely robust against surface diffusivity and backscattering, as was shown in Ref.~\cite{bib:miyawaki_higashitani_2018}. However, since the sign change in the order parameter in that case comes from reflected trajectories, the robustness might be lost at interfaces with finite transmission into other systems, e.g. in junctions \cite{bib:lofwander_shumeiko_wendin_2001}. In summary, the crucial factor for the phase to appear is a significant spectral weight of midgap states caused by sign-changing quasiparticle scattering trajectories.

\section{\label{sec:conclusion}Summary and conclusions}
The goal of this paper has been to provide a more complete picture of spontaneous symmetry-breaking tied to zero-energy Andreev states, and discuss experimental conditions where such phases can be observed. As an example, we have considered a particular phase with a spontaneous superfluid momentum due to pair-breaking interfaces in unconventional $d$-wave superconductors \cite{bib:hakansson_2015,bib:holmvall_2018_a,bib:holmvall_2018_b}. However, the results and the analysis presented in this paper can be extended to other phases and systems that host surface Andreev states, e.g. $p$-wave superconductors \cite{bib:suzuki_asano_2014,bib:dimitriev_2015,bib:zhelev_2016,bib:etter_bouhon_sigrist_2018,bib:miyawaki_higashitani_2018}.

In particular, we have studied how the realization of such phases is influenced by suppressing the spectral weight of the midgap states (via changing the angle $\theta$ between the pair-breaking interface and the $d$-wave crystal $\mathbf{\hat{a}}$-axis), by the sidelength $\mathcal{L}$ of the grain, as well as by surface roughness.

It was found that the transition temperature $T^{*}(\theta)$ into the symmetry-broken phase follows the angular dependence of the zero-energy state peak $N_{\mathrm{MGS}}(\theta)$, showing robustness against variations in $\theta$, even appearing at completely circular interfaces.

Furthermore, it was found that the sample-averaged observables (e.g. the heat-capacity jump in the phase transition) scale as $\mathcal{L}^{-1}$, down to a critical sidelength $\mathcal{L} = \mathcal{L}_{\mathrm{c}}$. At this sidelength, superconductivity starts becoming suppressed. Hence, sample-averaged observables are generally maximized at $\mathcal{L}_{\mathrm{c}}$. The critical sidelength depends on the shape of the sample, e.g. it was found that $\mathcal{L}_{\mathrm{c}} \approx 30\xi_0$ for a square grain.

With the above quantitative knowledge about how the shape and the size of the grain influence the symmetry-broken phases, grains with different degrees of mesoscopic surface roughness were analyzed. The conclusion was that any pair-breaking interface can generate spontaneous superflow, as long as the interface is within the critical angle and there is enough area around the interface to form the associated spontaneous currents. Finally, we discussed atomic surface roughness, referring to interfaces with diffuse quasiparticle scattering. Due to the results of Refs.~\cite{bib:higashitani_miyawaki_2015,bib:miyawaki_higashitani_2018}, the translational symmetry-breaking phase is expected to survive considerable atomic surface roughness, but more research is required.

In conclusion, any effect that broadens or reduces the spectral weight of zero-energy states will impede the realization of the symmetry-broken phases and the formation of the spontaneous superfluid momentum with associated magnetic flux. The advice to experimentalists aiming to study these phases is therefore to use systems with a maximized spectral weight of zero-energy Andreev states, with minimal interference from effects that broaden these states (e.g. atomic-scale surface roughness, impurities, strong external fields). For measurements of sample-averaged (e.g. thermodynamic) quantities, it is desirable to maximize the pair-breaking surface-to-volume ratio, as long as the volume does not become so small that superconductivity is severely suppressed. Thus, a specific suggestion would be to use heavy ion-bombardment to induce well-defined pair-breaking channels \cite{bib:walter_1998}. Another suggestion would be to deposit on a substrate a large array of rectangular or square-shaped thin-film $d$-wave grains with maximally pair-breaking edges, where the smallest sidelength is $\mathcal{L} = \mathcal{L}_{\mathrm{c}} \approx 30\xi_0$, and then look for either a heat capacity jump at $T=T^{*}$ with nanocalorimetry \cite{bib:diao_rydh_2016}, or the mesoscopic currents and flux that we previously reported on \cite{bib:hakansson_2015,bib:holmvall_2018_a,bib:holmvall_2018_b} with local probes, e.g. single-spin detectors \cite{bib:rugar_2004}, scanning-tunneling spectroscopy \cite{bib:nishio_2008,bib:cren_2011}, nano-SQUIDS \cite{bib:vasyukov_2013}, magnetometry \cite{bib:szechnyi_2017} and diamond cantilevers \cite{bib:pelliccione_2016,bib:ariyaratne_2018}.

There are still open questions regarding the survival of these symmetry-breaking phases at semi-transparent or transparent interfaces, and how they are influenced by quantum-size effects using fully microscopic theories \cite{bib:shanenko_peeters_2006,bib:zhang_peeters_2012,bib:zhang_peeters_2013,bib:huang_2016,bib:zha_2016,bib:yang_2018,bib:nagai_ota_tanaka_2017}. Furthermore, it would be interesting to see how the translational symmetry-breaking phase survives diffuse surface scattering \cite{bib:higashitani_miyawaki_2015,bib:miyawaki_higashitani_2018}, impurity effects \cite{PhysRevB.59.7102}, and in $p$-wave systems \cite{bib:suzuki_asano_2014,bib:dimitriev_2015,bib:zhelev_2016,bib:etter_bouhon_sigrist_2018,bib:miyawaki_higashitani_2018}.

\begin{acknowledgments}
We thank the Swedish Research Council for financial support. It is a pleasure to thank Mikael H\r{a}kansson for valuable discussions.
\end{acknowledgments}


%

\end{document}